# AV-TLX for measuring (mental) workload while driving AVs: Born from NASA-TLX but developed for the era of automated vehicles

Saeedeh Mosaferchi[1][0000-0001-9400-7198], Alireza Mortezapour[2*][0000-0001-6356-2244], Magnus Liebherr[3][0000-0001-9058-8079], Francesco Villecco[1][0000-0001-6545-4589], Alessandro Naddeo[1][0000-0001-7728-4046]

[1]Department of Industrial Engineering, University of Salerno, Via Giovanni Paolo II, 132, 84084 Fisciano, SA, Italy

[2]Department of Computer Science, University of Salerno, 84084 Fisciano, Italy

amortezapoursoufiani@unisa.it / amortezapour258@gmail.com

[3]Department of General Psychology: Cognition, University Duisburg-Essen, Duisburg, Germany

**Abstract**

The introduction of automated vehicles has redefined the level of interaction between the driver and the vehicle, introducing new tasks and so impose different workloads. Existing tools such as NASA-TLX and DALI are still used to assess driving workload in automated vehicles, despite not accounting for new tasks. This study introduces AV-TLX, a specialized tool for measuring workload in Level 3 automated driving. The development process began with a narrative literature review to identify the primary factors influencing workload. This was followed by a series of qualitative sessions during which the dimensions—and later the questions—of the questionnaire were designed. The tool's validity was first assessed using CVR and CVI indices, and its reliability and convergent validity were evaluated using a dynamic driving simulator with high fidelity. The final version of AV-TLX comprises 19 questions across 8 subscales, demonstrating excellent reliability (0.86) and validity (CVR > 0.78). An agreement scores between the results of AV-TLX and NASA-TLX in the simulation study was 0.6, which is considered acceptable for the consistency of two questionnaires. Furthermore, this questionnaire can be utilized in two ways: by reporting the overall workload and/or divided into 8 primary subscales, or by categorizing the questions into two groups—"takeover task" workload and "automated driving task" workload. The final version of this questionnaire, as presented in the paper, is available for use in future studies focusing on Level 3 automated driving.

## 1. Introduction

The mental workload of drivers stands out as a crucial factor shaping their driving performance [1]. By definition, the mental workload of a driver denotes how extensively they use their available cognitive resources behind the wheel and it represents the amount of cognitive capacity assigned for performing a driving task [2, 3]. Although this concept has been introduced beyond the context of driving, its evaluation within driving contexts spans over four decades [4]. So far, various subjective and objective methods have been devised to assess mental workload [5]. Nevertheless, given the subjective nature of mental resource utilization, questionnaires remain indispensable for assessing mental workload [6].

The main subjective instrument devised for assessing this concept is National Aeronautics and Space Administration-Task Load Index (NASA-TLX), introduced in 1986 [7]. Despite its presentation decades ago, this tool remains a cornerstone in assessing (mental) workload in different tasks including driving [8]. Notably, efforts have been made to develop more specialized tools for driving tasks, such as Driving Activity Load Index (DALI), which is based on the NASA-TLX [2]. Alongside these two methods, NASA-TLX being highly appreciated and DALI being specialized, other tools - such as SWAT and BEDFORD scales - have been employed to evaluate drivers' mental workload [9]. However, due to their limited usage and lack of specificity for driving, they are not further considered here.

NASA-TLX operates as a multidimensional approach, assessing drivers' workload across six distinct dimensions: Physical load, Mental load, Temporal load, Performance, Effort, and Frustration [10]. Extensive efforts were done to adapt this tool to various professions, environments and instruments such as Surgeon-TLX (for surgery context) and SIM-TLX (for virtual reality environments) which were developed [11, 12].

Inspired from NASA-TLX and tailored explicitly for driving tasks, the DALI tool has emerged. DALI delves into drivers' mental workload through six dimensions: Effort of attention, Visual demand, Auditory demand, Temporal demand, Interference, and Situational stress. Despite modifications to the subscales, calculation of its final score is similar to the raw score of NASA-TLX, calculating the mean of all sub-dimensions [13]. While this questionnaire was developed specifically for driving tasks and took remarkable steps by incorporating crucial parameters affecting mental workload in driving, it has a fundamental limitation these days. Given its introduction in the years before 2008, DALI fails to capture the era of automation.

These days, with technology advancing so quickly, the emergence of automated vehicles has revolutionized the concept of driving entirely [14]. There's even speculation that, over time, occupants of automated vehicles will transition from drivers to mere passengers [15]. This shift signifies that as automation in driving advances towards fully automated vehicles (L5) [16], numerous driving-related responsibilities will gradually shift from drivers to technology. Regardless of the profound evolution in the role of driving, the matter of drivers' mental workload and its assessment remains an open subject [17]. In recent years, numerous researchers have endeavoured to evaluate it across different levels of automated vehicles. While some researchers argue that automated driving can decrease drivers' workload, other studies counter that claim. They suggest that the abundance of information or infotainment available in automated vehicles, owing to the plethora of interfaces and the information explosion era, may actually heighten drivers' mental workload [18, 19]. Despite the advent of automated vehicles, the shifting landscape of driver responsibilities, and the persistent inquiry into the impact of these factors on drivers' mental workload, the tools, and questionnaires for assessing mental workload have not kept pace with this rapid evolution. This might explain why established tools such as NASA-TLX and DALI continue to be used even in this era.

By shedding light on the shortcomings of these tools, particularly their outdated nature in the era of automated vehicles, the objective of the present study was to identify crucial parameters influencing drivers' mental workload from previous literature. These parameters then were augmented with qualitative insights of the experts in the field and synthesized to develop a new tool, Automated Driving-Task load index (AV-TLX), tailored to the context of level 3 of automated driving.

## 2. Material and Methods

To develop an updated tool for assessing the (mental) workload of drivers or users of Level 3 automated vehicles, we employed a mixed-method study design [20]. The process began with a narrative review and ended in an experimental study (following sections). The study was performed in accordance with the ethical standards laid down in the Declaration of Helsinki and approved by the local ethics committee at the University of Salerno, Italy. All par-ticipants provided written informed consent prior to the experiment and were informed that they could end participation at any time without reprisal.

### 1.1. Conducting a narrative review to extract the influential parameters

In the previous section, we highlighted major concerns about how existing tools are falling behind as technology advances. Consequently, the existing tools for measuring workload lack a comprehensive list of parameters. To try to bridge this gap, we conducted a narrative review to identify and synthesize those parameters that influence the mental workload of drivers in automated vehicles. We opted for a narrative review over a systematic review to avoid limiting ourselves to a narrow research question [21]. This approach allowed us to shed light on the previous literature and gather a broad spectrum of parameters pertaining to the concept of (mental) workload in automated vehicles. To achieve this objective, we compiled keywords associated with (mental) workload and automated vehicles and conducted searches exclusively in the Scopus and Web of Science. Additionally, we completed our search by reviewing the first 10 pages of Google Scholar results. The articles identified through this process underwent initial screening by one of the authors, who evaluated them based on their titles and abstracts. Unlike systematic review studies, we employed general criteria to encompass any potential parameters relevant to (mental) workload, cognitive load and/or strain, and similar concepts in the context of automated vehicles. Articles deemed highly relevant were subjected to full-text review. Following the identification of the initial list of parameters, we proceeded to classify them in the following step.

### 1.2. Considering previous tools for categorizing the parameters

Two primary approaches for categorizing the parameters are well documented and common in the literature. The first approach involved clustering the identified parameters based on established models found in previously published literature, such as cognitive load theory and its intrinsic and extrinsic components [22]. The second approach was to utilize similar tools, even in the absence of theoretical models. At this juncture, we opted for the second method. This decision stemmed from the fact that despite lacking updates in previous tools such as NASA-TLX and DALI, these tools remained highly appreciated and had been extensively utilized in numerous studies. As previously mentioned, both tools assess (mental) workload through six subscales. While some of these subscales overlap, others differ. We attempted to utilize a combination of these items for their initial classification. However, given the lack of updates in these tools regarding technological advancements and the inclusion of new parameters in our list, new subscales were added to complete the existing tools. Our initial tool has 9 different subscales.

### 1.3. Finalizing the parameters and re-arrangement of the main categories in a series of qualitative study

Over three sessions, each lasting between 1-2 hours, we discussed the initial parameters and their categorization with a group of experts in this field. These experts voluntarily accepted our invitation via LinkedIn or email to participate in the qualitative phase of the study. A total of seven different experts took part in these sessions. The outcome of these discussions was a reduction in the number of categories from nine to eight, along with the merging of some parameters, omitting/adding some others. The final agreed-upon model after the qualitative sessions is presented in Figure 1 (supplementary section). Since this study focuses on Level 3 automated driving, concepts such as take-over remain important in AV-TLX [17].

### 1.4. Making questions for each parameter and each dimension

Appropriate questions were designed for the remaining subscales and dimensions in the context of automated vehicles. To achieve this, relevant tools such as NASA-TLX and DALI were reviewed, and principles of questionnaire design for human studies were also considered [23]. The initial version of the questionnaire consisted of 23 questions. A 7-point Likert scale, ranging from "Strongly Disagree" to "Strongly Agree," was designated for AV-TLX.

### 1.5. Conducting a CVR, CVI session

As part of the validation process the CVR (Content Validity Ratio) and CVI (Content Validity Index) methods were used [24, 25]. This approach which is normally employed in the validation of new questionnaires, involves collecting expert opinions on the content validity of the questions. Content validity was evaluated across four key dimensions: Necessity, Simplicity, Relevance and Specificity, Transparency and Clarity. Based on the number of participating experts, the minimum acceptable scores for CVR and CVI indices were determined. At this stage, nine different experts provided their feedback on these four dimensions. As a result, a revised version of the questionnaire was developed with slight modifications, consisting of 19 questions across the same eight dimensions as before (CVR > 0.78).

The final version of the questionnaire is presented in Table 1 (supplementary section). This version, which demonstrated also acceptable reliability according to Cronbach's alpha method (using the driving simulator which is presented in the next phase), is suitable for use in future studies. Considering 7-point Likert scale, the minimum score that can be obtained is 19 (19×1), and the maximum score is 133 (19×7). It is essential to ensure that the questions' orientations are standardized when using this questionnaire (to be all positive). For the application of this instrument in future studies, two primary categories have been foreseen by our team: assessing workload during take-over and assessing workload during automated driving. To accommodate this, 19 questions have been labeled as TO and AD. Worth noting that using only one section of the questions is possible. In this case, the total score for that specific section should be calculated and applied based on the same 7-point Likert scale.

As shown in Table 1 (supplementary section), all 19 questions are categorized into 8 dimensions. A separate categorization has been also considered for automated driving (AD) and/or take-over (TO) tasks.

### 1.6. Using the questionnaire in a pilot experimental study

Participants:

Twenty-five (16M, 9F) students at the University of Salerno participated voluntarily in the experiment. All participants had driving license and were Italian to control for the potential confounding effect of nationality, including differences in technological exposure and experience with automated vehicles.

Apparatus:

A medium to high fidelity dynamic driving simulator was used (Figure 2). The system used three 65-inch 1080 screens, delivering a 120° field of view. The simulator was equipped with an adjustable seat and backrest, a steering wheel, accelerator and brake pedals, as well as an immersive surround sound system. The driving experience was powered by BeamNG software.

Procedure:

First, the researchers explained the objectives and procedure of the study to the participants. If they agreed to take part, they signed a consent form. Subsequently, they completed a demographic questionnaire. Following this, they proceeded to sit inside the driving simulator. Participants adjusted the seat to their preferred comfort level before the test began. The experiment consisted of six stages, lasting a total of 15 minutes. These six stages were divided into two distinct driving modes: three were manual, and three were automated. During the automated phases, participants were instructed to solve a Rubik's cube puzzle using a 12-inch Samsung tablet (as a NDRT). The sequence of the stages was as follows:

1. Fully Automated (3 min, daytime): Participants drove on a high-traffic highway in sunny conditions. Thirty seconds before switching to manual, they received two warnings, followed by a vocal alarm for takeover (***Attenzione. Il controllo della guida ti verrà restituito entro 30 secondi. Quando senti il suono del campanello, I comandi dell'auto sono stati disattivati***).

2. Manual Driving (2 min, daytime): Participants drove manually under the same conditions. Fifteen seconds before automation resumed, they received a vocal warning to return control to the system (*Attenzione. Dopo 15 secondi l'auto tornerà in modalità di guida autonoma*).

3. Automated (3 min, nighttime): Participants engaged with a tablet while driving on a congested highway at night. Five seconds before the takeover, a vocal alarm was given (***Attenzione. In 5 secondi avri il controllo della guida. Quando senti il suono del campanello, I comandi dell'auto sono stati disattivati***).

4. Manual Driving (1 min, nighttime, tunnel): Participants drove through a crowded tunnel at night. Ten seconds before switching back to automation, a vocal warning was issued (***Attenzione. L'auto tornerà alla modalità di guida autonoma in 10 secondi***).

5. Fully Automated (4 min, nighttime, fog): Driving included tunnels, busy roads, and a mountainous area with heavy fog and traffic. Participants used the tablet, receiving a takeover warning one minute before manual driving (***Attenzione. Ricevera il controllo della duida in 1 minuto***).

6. Manual Driving (2 min, nighttime, fog): The last stage mirrored previous conditions—nighttime, foggy, and high traffic—before concluding the experiment.

After completing the simulation, participants left the simulator and sat comfortably on a chair while responding to AV-TLX. They also answered a question related to motion sickness and subsequently completed the NASA-TLX questionnaire.

### 1.7. Checking the reliability of the AV-TLX

After the experiment, the reliability score was calculated for 25 participants by Cronbach alpha method. This method is recommended in previous researches for Likert-based questionnaires [26]. The obtained alpha score (= 0.86) is considered totally acceptable. In addition to the acceptable validity discussed in the previous section, this instrument is suitable for use in future studies due to its excellent reliability.

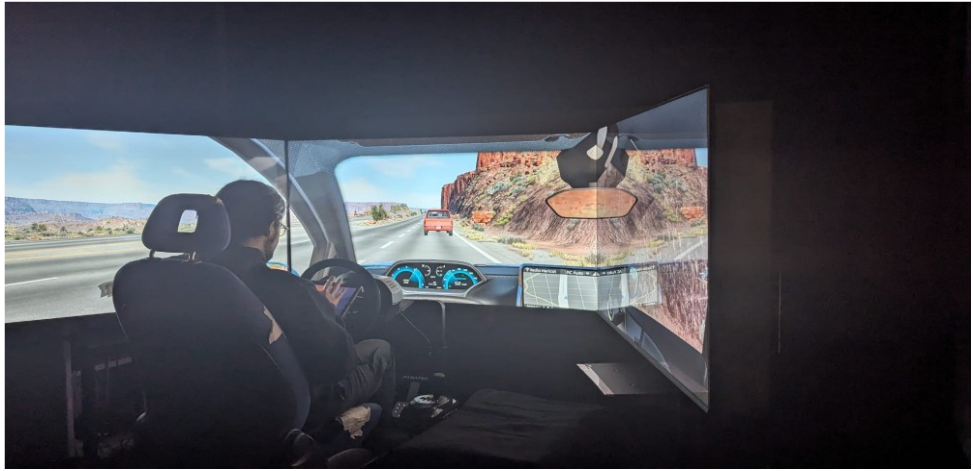

Figure 1. Driving simulator environment

# 3. Results

The participants had a mean age of 27 years (SD=3.3). Among them, one was an undergraduate student, 13 were master's students, and 11 were Ph.D. students. Their average level of familiarity with automated vehicles was 4.28 (SD=1.3) out of 7, with only two participants having prior experience using automated vehicles in real roads. Ten attendees had previously participated in studies involving driving simulators. One reported a history of cognitive issues that had been resolved and did not affect his performance in the study. After the simulator session, 2 participants reported mild motion sickness; however, none of these cases led to the termination of the experiment. Participants had the option to withdraw from the study in case of any issues, including motion sickness. Table 2 presents the workload results of the participants, on each subscale as well as the overall workload. As presented, both AV-TLX and NASA-TLX for the participants were in the mid-range (AV-TLX: 3.17 out of 7 (similar to 45 out of 100) vs. NASA-TLX: 41.8 out of 100).

**Table 1.** (Mental) Workload score of the participants

| AV-TLX (out of 7) | Mean (SD) | NASA-TLX (out of 100) | Mean (SD) |
|---|---|---|---|
| AV-TLX (Physical) | 2.96 (1.47) | NASA-TLX (Mental) | 49.4 (24.2) |
| AV-TLX (Frustration) | 3.2 (1.25) | NASA-TLX (Physical) | 32.4 (23.3) |
| AV-TLX (Personal Performance) | 3.16 (1.44) | NASA-TLX (Temporal) | 38 (22) |
| AV-TLX (Contextual stress) | 3.3 (0.75) | NASA-TLX (Performance) | 47 (22.9) |
| AV-TLX (Temporal Demand) | 2.8 (1.52) | NASA-TLX (Effort) | 50.8 (26) |
| AV-TLX (Safety Concerns) | 3.12 (1.14) | NASA-TLX (Frustration) | 33.8 (25) |
| AV-TLX (High-level Interaction) | 3.24 (0.98) | NASA-TLX (Total Raw) | 41.8 (17) |
| AV-TLX(Visual/Vocal/Tactile Demands) | 3.29 (0.94) | | |
| AV-TLX (Total) | 3.17 (0.78) | | |

To assess the agreement between AV-TLX and NASA-TLX to ensure the convergent validity of AV-TLX, Pearson's correlation test was used, as both variables and their subscales are quantitative (and most of them had normal distribution based on Shapiro-Wilk test). The results are presented in Table 3.

**Table 2.** Agreement scores of two questionnaires and the subdimensions by bivariate analysis (**Bolds are significant**)

| Pearson coefficient (P-value) | NASA-TLX (Total) | NASA-TLX (Mental) | NASA-TLX (Physical) | NASA-TLX (Temporal) | NASA-TLX (Performance) | NASA-TLX (Effort) | NASA-TLX (Frustration) |
|---|---|---|---|---|---|---|---|
| AV-TLX (Total) | **0.6 (0.001)** | | | | | | |
| AV-TLX (Physical) | | | **0.58 (0.002)** | | | | |
| AV-TLX (Frustration) | | | | | | | **0.59 (0.002)** |
| AV-TLX (Personal Performance) | | | | | 0.001 (0.99) | | |
| AV-TLX (Contextual stress) | | | | | | | 0.33 (0.09) |
| AV-TLX (Temporal Demand) | | | | 0.004 (0.81) | | | |
| AV-TLX (Safety Concerns) | | | | | | | |
| AV-TLX (High-level Interaction) | | | | | | | |
| AV-TLX (Visual/Vocal/Tactile Demands) | | | | | | | |

The agreement results presented in Table 3 indicate that there is an acceptable and significant agreement between the overall scores of the two instruments. Additionally, the agreement between the corresponding subscales of each instrument ranges from moderate to good. However, no agreement was observed in the subscales of Temporal Demand and (Personal) Performance between the two instruments. There was a good agreement between the two instruments, which clearly indicates that they are aligned in measuring a common construct (Workload). In fact, AV-TLX demonstrates acceptable convergent validity. To support Table 2 and compare the overall scores of the two instruments, Table 3 also proves this agreement.

## 4. Discussion

With the introduction of (semi-)automated vehicles, the concept of driving and the driver's role has fundamentally changed, shifting from active control to passive supervision. This transformation, along with evolving interactions between the driver, vehicle, road, and environment, has significantly redefined driving workload [25]. However, it remains an open topic in the field of human factors engineering [17]. The aim of the present study was to introduce a new tool (AV-TLX) designed for the emerging context in Level 3 automated driving.

### 4.1. New context-based adopted versions developed from NASA-TLX

As mentioned before, NASA-TLX was introduced over 40 years ago in the aviation/aerospace domain [4]. In recent years, various researchers have attempted to develop new versions of this tool tailored to specific applications and contexts, reporting their validity and reliability. For instance, Mark R. Wilson et al. attempted to adapt NASA-TLX for the surgical profession [11]. This new tool (Surgeon-TLX), introduced in 2011, retained six dimensions but incorporated modifications to better align with the demands of surgical tasks. Similarly, a version designed for assessing workload in prosthesis users, known as Prosthesis Task Load Index (PROS-TLX), has been developed [27]. One of the most recent efforts to update NASA-TLX is the introduction of simulation task load index (SIM-TLX) in 2019, which was developed to address the specific requirements of virtual reality environments [12]. A key distinguishing feature of SIM-TLX compared to previous adaptations of NASA-TLX is the expansion of its dimensions from six to ten. In their study, the authors declared that the six dimensions of the original NASA-TLX were not sufficient to fully capture the workload demands of these emerging virtual environments, necessitating the addition of new dimensions.

If we consider newer versions of NASA-TLX specifically in the driving domain, the best example is DALI, which was introduced in 2008 [13, 28]. This tool has been completely explained in the introduction section. However, since nearly two decades have passed since its introduction, and given the significant changes in driving tasks brought about by automated driving, this study aims to introduce AV-TLX specifically designed for Level 3 automated vehicles. This tool addresses key tasks such as takeover scenarios, automated driving operations, and the reduced responsibilities of the driver. AV-TLX consists of eight main subscales and 19 questions. Some of the subscales, with slight modifications, were also present in previous versions of NASA-TLX and DALI, such as physical demands and visual demands. However, others have been introduced specifically due to the fundamental shift in driving tasks in the era of automated driving. Notable examples include high-level interactions (explainability) and even new questions in temporal demand (time-budget in take-over) or personal performance (NDRT performance) subscales, which reflect the unique demands of Level 3 automated vehicle operations [29]. As previously mentioned in the example of SIM-TLX [12], increasing the number of dimensions from six in the original NASA-TLX—and even in DALI—to eight in our newly developed tool was essential.

## 4.2. Comparing RTLX and AV-TLX scores with previous studies

In the scenario conducted for the validity and reliability assessment of this new tool, driving workload in a mixed condition—where manual and automated driving were combined, including takeover transitions—was approximately 3.2 out of 7 in AV-TLX and 42 out of 100 in the original NASA-TLX. This level indicates a low to moderate workload in the mixed task of automated and manual driving, which is consistent with some previous studies. However, there are other studies that have reported either lower or higher values. These differences are fully understandable, as individual factors play a role in each specific scenario in every study.

Jork Stapel and his colleagues compared the workload of experienced drivers using the automated mode of a Tesla with inexperienced drivers who had no prior exposure to automated driving [30]. Since most participants in our study had no prior experience with automated driving (23 out of 25), comparing these two similar conditions is logical. In the study by Jork Stapel et al., the driving workload (using NASA-TLX) of inexperienced drivers was reported as 42.60±17.1 for manual driving and 42.90±20.1 for automated driving in a complex road scenario, which is similar to our road geometry. It is evident that the average of these two values, representing the mixed condition, also indicates approximately 42 out of 100, which aligns with the findings of our study. A key difference between these two studies is the use of a simulated driving scenario in our study, whereas the other study utilized a real-world driving scenario with a Tesla vehicle on regular roads in the Netherlands.

Researchers from the United Kingdom compared the mental workload scores of participants in real-world conditions across three different automated vehicles with Level 2 automation [31]. In the complex scenario, which imposed a workload similar to the present study, participants reported a NASA-TLX workload score of 40–50 across the three different automated vehicles with Level 2 automation for the mixed condition of manual and automated driving. In a relatively similar scenario conducted in Canada, a Tesla ( Level 2) was examined in both manual and automated modes on real roads [32]. The average NASA-TLX score was 7.59/21, roughly 38/100 on a standardized scale. However, a key difference between the studies mentioned and the present work lies in the automation level (Level 2 vs. Level 3). Another notable difference is the contrast between driving in a simulator and real-world driving conditions. After reviewing three studies in real-world driving conditions, a driving simulator study showed that in a mixed driving condition between manual and automated modes, drivers reported a NASA-TLX workload score of 43.30±14.93 [33]. Also, this result is consistent with the current study's result. In another similar research on the combination of manual and Level 3 automated driving, the workload score was approximately 40 (around 30 for automated driving and 55 for manual driving) [34]. Overall, it can be concluded that the combined workload score for automated and manual driving (both in simulators and real-world studies) has been reported at approximately 40-50 out of 100 in a significant number of studies. This aligns with our findings, suggesting that the newly introduced tool (AV-TLX) can effectively quantify driver workload in the era of automated vehicles, just similar to NASA-TLX (as the best documented tool).

### 4.3. Take-over (TO) questions of AV-TLX in comparison to take-over perceived workload in previous studies

In AV-TLX, 7 out of the 19 questions are specifically dedicated to measuring workload during the takeover task in transition between manual and automated mode. The workload score of 25 participants on these 7 questions, reflecting their workload during the takeover process, was 3.2 ± 0.89 (on a 7-point scale), approximately equivalent to 45 out of 100. The minimum and maximum perceived scores were 2.14 and 5.86, respectively (Range: 3.72). In general, various parameters have been considered influential in modeling driver take-over performance by researchers. These include time budget, traffic density, non-driving-related tasks, and repetition [35], all of which plus other ones are covered in the seven questions of the questionnaire designed for this task. In a recent study, the workload of 19 drivers was reported in a driving simulation study at Level 3 automation across five different driving conditions. The workload of drivers during the take-over task after an automated driving phase at Level 3 was reported as 12.48 out of 21 [36]. This level is slightly higher than the workload reported in the present study (≈45 out of 100 vs. ≈59 out of 100). In another study, where participants were using their smartphones before performing a takeover, the reported takeover workload was 49 on the 0–100 NASA-TLX scale, which is similar to the results of the present study [37]. It appears that dividing the AV-TLX questions into two main subsets -one for automated driving tasks (AD in the AV-TLX) and one for takeover tasks (TO in AV-TLX)- yields acceptable and comparable results with previous studies. Therefore, this tool can be effectively used with this two-state categorization (instead of 8 subdimensions and one overall workload score).

### 4.4. Pair-wise comparison of convergent validity of AV-TLX with NASA-TLX

As mentioned earlier, some dimensions of AV-TLX were derived from previous questionnaires (NASA-TLX and/or DALI), while others are entirely new, designed specifically to address the new generation interactions between drivers and L3 automated vehicles. The optimal approach for convergent validation in this context is to compare the corresponding subscales of the two tools pairwise, as well as their overall workload scores. Six pairwise comparisons for this purpose are reported in Table 3, of which three comparisons showed a highly significant correlation, while the other three comparisons revealed no significant correlation between the two variables.

The temporal demand subscale in both tools showed no significant correlation. Upon examining the questions asked in each (time pressure in general driving task vs. time-budget for only take-over task), it becomes evident that the lack of correlation between these two subscales is logical, as they assess two distinct cognitive dimensions. Compared to the overall driving task with a variety of sub-tasks present in manual vehicles, takeover has been defined as a new interaction for driving in semi-automated vehicles [37]. Therefore, in the NASA-TLX, participants likely tried to rate the entire driving task, whereas in the AV-TLX, due to the specific focus on the takeover task, they answered only to take-over task.

This issue can also be extended to the relationship between the "Performance" subscale from NASA-TLX and the "Personal Performance" subscale from AV-TLX. The questions in this dimension of NASA-TLX still relate to the individual's performance in the entire driving process [10]. However, in AV-TLX, the subscale concerning Personal Performance focuses on two distinct aspects: one related to the individual's performance in Non-Driving Related Task (NDRT) and the other on performance in executing a takeover request. These two subcategories did not exist in previous driving tasks of manual vehicles and, naturally, were not covered by NASA-TLX either [1, 38]. The importance of having adequate performance in NDRT has been highlighted as one of the key benefits of automated vehicles [39]. Since NDRT directly affects situational awareness and consequently has a significant impact on cognitive workload in driving, its inclusion in this questionnaire can be considered a strong advantage [40]. On the other hand, the three significant correlations between the two questionnaires—one in terms of the overall workload score and the other one in the Physical-Physical, and Frustration-Frustration sub-scales—demonstrate a considerable similarity between the two instruments.

Perhaps the most crucial pairwise comparison pertains to the overall workload score obtained from both tools. Given the agreement score of $r = 0.6$ between the two tools, it can be concluded that AV-TLX effectively measures driving workload similarly to NASA-TLX. Acceptable ranges for assessing agreement between two tools have been reported in previous studies, generally starting from around 0.5 and above. Yuan-mei Xiao et al. assessed the consistency of NASA-TLX and SWAT by Pearson correlation coefficient and reported a 0.49 consistency result as a good level [41]. In some studies, such as a 2004 study which is reported the agreement between NASA-TLX, SWAT, and WP, the agreement score between different workload assessment tools has occasionally been reported as high as 0.95 and more [42].

Regarding the Physical Demand sub-scale in AV-TLX, although it refers to a different process compared to the Physical Demand dimension in NASA-TLX, both finally will assess the physical strain on the driver's body while using the vehicle. This significant correlation between the two subscales can be attributed to the perceived physical strain caused by the driver's posture in the vehicle [43]. Similarly, this concept can be extended to the Frustration subscale in both tools. In NASA-TLX, the Frustration sub-scale assesses feeling annoyed and discouraged, while in AV-TLX, Frustration contains two specific questions address the perceived lack of a meaningful role in comparison to the vehicle and the excessive repetition of takeover requests and their annoying impact on the driver [44, 45].

The lack of reported pairwise correlations between the remaining subscales of the two tools suggests that the authors believed there was no logical connection between them. This assumption stems from the significant modifications in AV-TLX compared to NASA-TLX, which were driven by fundamental changes in driver interaction between automated and conventional vehicles. However, this issue could be further tested in future studies and different driving scenarios.

### 4.5. Limitations and Future directions

Despite advantages such as a high-fidelity driving simulator used in this study, certain limitations may have influenced our results. The most significant limitation is the small sample size of participants (N=25). Additionally, the lack of control over background variables affecting workload may have negatively impacted the findings. For future research, it is recommended to test AV-TLX in larger sample studies, across diverse driving scenarios in simulators, and even in real-world driving conditions.

## 5. Conclusion

Although the value of NASA-TLX has been well documented, AV-TLX has successfully provided a meaningful update for assessing workload in the era of automated driving. Its excellent validity and reliability, along with its specificity for Level 3 automated vehicles, make it a promising candidate for use in future studies. This new questionnaire can be utilized in two ways. One approach involves reporting the overall workload score and breaking it down into eight primary subscales. Alternatively, the questions can be categorized into two groups: takeover workload (7 out of 19 questions) and automated driving task workload (13 out of 19 questions). One question is repeated for both.

# Supplemetary section:

**Table 3.** Final version of AV-TLX ($^R$ shows the required reverse scoring for final calculation)

| Num | Dimension | Questions | Application |
|---|---|---|---|
| 1 | Physical Demand | Taking back manual control from automated driving, caused me physical discomfort $^R$. | TO |
| 2 | Physical Demand | The limited space for personal tasks during automated driving put physical pressure on my body $^R$. | AD |
| 3 | Frustration | Having a less significant role compared to the vehicle's advanced capabilities created mental load for me $^R$. | AD |
| 4 | Frustration | Several requests from the vehicle to take control, especially in simple situations, kept my mind under constant load $^R$. | TO |
| 5 | Personal Performance | While the car was driving automatedly, I could concentrate on my personal activities without any mental strain. | AD |
| 6 | Personal Performance | I took back control of the car as required, without feeling any mental strain. | TO |
| 7 | Contextual Stress | Adverse conditions like bad weather, low light, or complex road increased my mental load both during automated driving and take-over time $^R$. | AD/TO |
| 8 | Contextual Stress | During long journeys, using automated mode decreased my mental load. | AD |
| 9 | Contextual Stress | My mental load decreased as I trusted the cars' sensors to detect the environment. | AD |
| 10 | Temporal Demand | Having enough time to do take-over reduced my mental load. | TO |
| 11 | Visual/vocal/Tactile demand | The displays' size, location, and design increased mental load when using them $^R$. | AD |
| 12 | Visual/vocal/Tactile demand | The clarity and amount of information on the displays were mentally demanding $^R$. | AD |
| 13 | Visual/vocal/Tactile demand | The car's vocal interactions are well-designed making the journey mentally undemanding $^R$. | AD |
| 14 | Visual/vocal/Tactile demand | Effective tactile interactions with the car decreased my mental load. | AD |
| 15 | High-level perceptual demand | The car's transparency in making decisions reduced my mental load. | AD |
| 16 | High-level perceptual demand | The car's driving pattern which matched my preferences, reduced my mental load. | AD |
| 17 | Safety Concerns | Doing non-driving tasks in the limited space behind the wheel created a mental load for me $^R$. | AD |

| 18 | The emergency stop button gave me a sense of control and safe and reduced my mental load. | TO |
| 19 | Due to my low situational awareness, the take-over request caused significant mental load [R]. | TO |

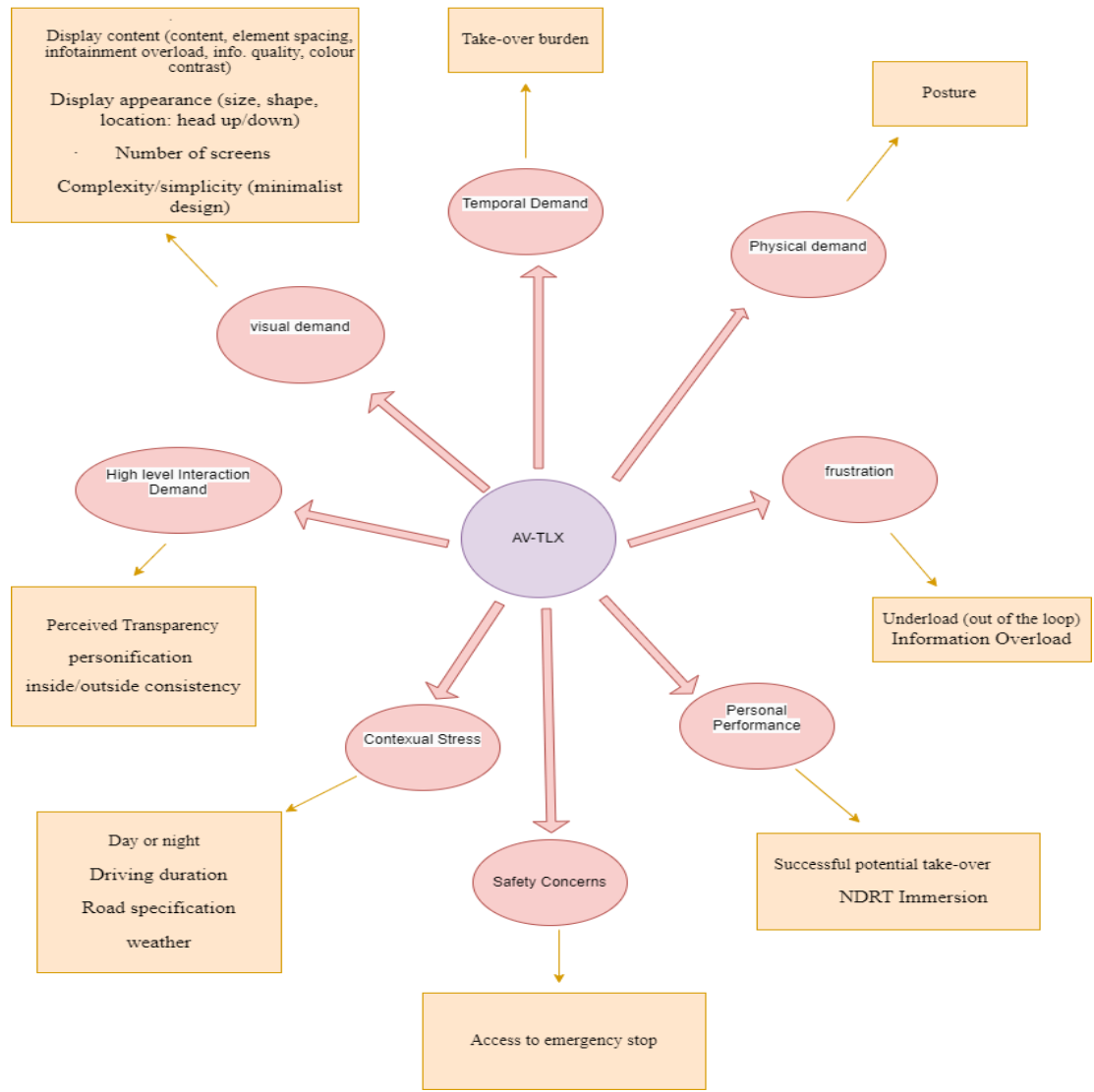

Figure 2. The agreed categorization of the parameters